\documentclass[aps,prc,preprint,showpacs,superscriptaddress,amsmath,amssymb]{revtex4}
\usepackage{graphicx}
\usepackage{dcolumn}
\usepackage{bm}
\usepackage{color}
\usepackage{booktabs, longtable}
\usepackage{multirow}
\usepackage{CJK}
\usepackage{ltxtable}
\usepackage{rotating}

\begin{document}
\begin{CJK*}{GBK}{song}
\title{Green's function method for single-particle resonant states in relativistic mean field theory}

\author{T. T. Sun}
\affiliation{School of Physics and State Key Laboratory of Nuclear Physics and Technology, Peking University, Beijing 100871, China}
\author{S. Q. Zhang}
\affiliation{School of Physics and State Key Laboratory of Nuclear Physics and Technology, Peking University, Beijing 100871, China}
\author{Y. Zhang}
\affiliation{Department of Physics, Faculty of Science, Tianjin University, Tianjin 300072, China}
\author{J. N. Hu}
\affiliation{School of Physics, Nankai University, Tianjin 300071, China}
\author{J. Meng}
 \affiliation{School of Physics and State Key Laboratory of Nuclear Physics and Technology, Peking University, Beijing 100871, China}
 \affiliation{School of Physics and Nuclear Energy Engineering, Beihang University, Beijing 100191, China}
 \affiliation{Department of Physics, University of Stellenbosch, Stellenbosch 7602, South Africa}

\date{\today}

\begin{abstract}
Relativistic mean field theory is formulated with the Green's function method in coordinate space to investigate the single-particle bound states and resonant states on the same footing. Taking the density of states for free particle as a reference, the energies and widths of single-particle resonant states are extracted from the density of states without any ambiguity. As an example, the energies and widths for single-neutron resonant states in $^{120}$Sn are compared with those obtained by the scattering phase-shift method, the analytic continuation in the coupling constant approach, the real stabilization method and the complex scaling method. Excellent agreements are found for the energies and widths of single-neutron resonant states.

\end{abstract}

%25.70.Ef Resonances
%21.10.Pc Single-particle levels and strength functions
%21.60.Jz Nuclear Density Functional Theory and extensions (includes Hartree-Fock and random-phase approximations)

\maketitle

\section{Introduction}\label{Chapter1}
With the development of the radioactivity ion beam facilities, the study of exotic nuclei with unusual $N/Z$ ratios has attracted world wide attention. Unexpected properties very different from those of normal nuclei have been observed, such as halo phenomena~\cite{PRL1985Tanihata55}, giant halo~\cite{PRL1998MengJ80}, new magic number~\cite{PRL2000OzawaA84}, and deformed halo as well as shape decoupling~\cite{PRC2010Zhou82}, etc. In exotic nuclei, the neutron or the proton Fermi surface is very close to the continuum threshold, thus the valence nucleons can be easily scattered to the single-particle resonant states in the continuum and the couplings between the bound states and the continuum become very important~\cite{PRC1996DobaczewskiJ53, PRL1996Meng77, PRL1997Poschl79, PRL1998MengJ80, PPNP2006MengJ}. For example, the self-consistent relativistic continuum Hartree-Bogoliubov (RCHB) calculations suggested that the neutron halo in $^{11}{\rm Li}$ is formed by scattering Cooper pairs to the $2s_{1/2}$ level in the continuum~\cite{PRL1996Meng77} and predicted giant halos in exotic Zr and Ca isotopes, which are formed with more than two valence neutrons scattered as Cooper pairs to the continuum~\cite{PRL1998MengJ80, PRC2002MengJ65, SciChinaSerG2003ZhangSQ46}. This novel giant halo phenomenon was further verified by non-relativistic density functionals~\cite{PRC2006TerasakiJ74, PRC2006Grasso74, PRC2012ZhangY86}. It should be noted that, by including only the contribution of the resonant states, the giant halos were also reproduced by the relativistic mean-field calculations with pairing treated by the BCS method~\cite{PRC2003Sandulescu68}. Therefore, the properties of the resonant states close to the continuum threshold are essential for the investigation of exotic nuclei.

Based on the conventional scattering theories, many approaches, such as $R$-matrix theory~\cite{PhysRev1947Wigner72, PRL1987Hale59}, $K$-matrix theory~\cite{PRC1991Humblet44}, $S$-matrix theory~\cite{Book1972Taylor-ScatteringTheor, PRC2002CaoLG66}, and Jost function approach~\cite{PRL2012LuBN109, PRC2013LuBN88}, have been developed to study the single-particle resonant states. Meanwhile, the techniques for bound states have been extended for the single-particle resonant states, such as, the analytic continuation in the coupling constant (ACCC) approach~\cite{Book1989Kukulin-TheorResonance, PRC1997Tanaka56, PRC1999Tanaka59, PRC2000Cattapan61}, the real stabilization method (RSM)~\cite{PRA1970Hazi1, PRA1971Fels4, PRA1972Fels5, PRA1976Taylor14, PRL1993Mandelshtam70, PRA1994Mandelshtam50, PRA1999Kruppa59}, and the complex scaling method (CSM)~\cite{PhysRep1983Ho99, PRC1986Gyarmati34, PRC1988KruppaAT37, PRL1997Kruppa79, PRC2006Arai74, CPC2010Guo181}.

Combining with the relativistic mean field (RMF) theory which has achieved great successes in describing both the stable and exotic nuclei~\cite{ANP1986Serot, RepProgPhys1989Reinhard52, PPNP1996Ring, PhysRep2005Vretenar409, PPNP2006MengJ}, some of the above methods for single-particle resonant states have been introduced to investigate the resonances. As examples, the RMF-ACCC approach is used to give the energies and widths~\cite{CPL2001YangSC18} as well as the wave functions~\cite{PRC2004ZhangSS70, EPJA2007ZhangSS} of resonant states. Similar applications for the Dirac equations with square well, harmonic oscillator and Woods-Saxon potentials can be seen in Ref.~\cite{CPL2004ZhangSS21}.
The RMF-RSM approach is introduced to study the single-particle resonant states in $^{120}{\rm Sn}$~\cite{PRC2008ZhangL77}. The RMF-CSM is developed to describe the single-particle resonant states in spherical~\cite{PRC2010GuoJY82, PRC2014Zhu89} and deformed nuclei~\cite{PRC2012LiuQ86, PRC2014Shi90}. The single-particle resonant states in deformed nuclei have been investigated by the coupled-channel approach based on the scattering phase-shift method as well~\cite{PRC2010LiZP81}.

Green's function method~\cite{Book2006Eleftherios-GF} is also an efficient tool for single-particle resonant states. Non-relativistically and relativistically, there are already many applications  of the Green's function method in nuclear physics to study the contribution of continuum to the ground states and excited states. In 1987, Belyaev \emph{et al.} constructed the Green's function in the Hartree-Fock-Bogoliubov (HFB) theory in the coordinate representation~\cite{YadFiz1987Belyaev45}. In Ref.~\cite{NPA2001Matsuo696}, Matsuo applied this Green's function in the quasi-particle random-phase approximation (QRPA), with which the collective excitations coupled to continuum states can be described~\cite{PTPS2002Matsuo146, PRC2005Matsuo71, NPA2007Matsuo788, PTP2009Serizawa, PRC2009Mizuyama79, PRC2010Matsuo82, PRC2011Shimoyama84}. In Ref.~\cite{PRC2009Oba80}, Oba \emph{et al.} extended the continuum HFB theory with the Green's function method for deformed nuclei. In Ref.~\cite{PRC2011ZhangY83}, Zhang \emph{et al.} developed the fully self-consistent continuum Skyrme-HFB theory with Green's function method, which is further extended for odd-$A$ nuclei~\cite{arXivSun2014}. Relativistically, based on the Green's function of the Dirac equation~\cite{PRB1992Tamura45}, the relativistic continuum random-phase-approximation (RCRPA) was developed to study the contribution of the continuum to nuclear excitations in Refs.~\cite{PRC2009DaoutidisRing80, PRC2010DingY82}. The advantages of Green's function method include treating the discrete bound states and the continuum on the same footing, providing both the energies and widths for the resonant states directly, and having the correct asymptotic behaviors for the wave functions.

In this work, we formulate the RMF theory with the Green's function method (RMF-GF) in coordinate space to investigate the single-particle resonant states. Taking the nucleus $^{120}$Sn as an example, the energies and widths for neutron resonant states will be obtained from the density of states calculated by the Dirac Green's function, and compared with those by the RMF-S method~\cite{PRC2002CaoLG66}, the RMF-ACCC approach~\cite{PRC2004ZhangSS70}, the RMF-RSM~\cite{PRC2008ZhangL77}, and the RMF-CSM~\cite{PRC2010GuoJY82}. In Sec.~\ref{Chapter2}, we give briefly the formulations of the RMF theory and the Green's function method. Numerical details are presented in Sec.~\ref{Chapter3}. After the results and discussions in Sec.~\ref{Chapter4}, a brief summary is drawn in Sec.~\ref{Chapter5}.

\section{THEORETICAL FRAMEWORK}\label{Chapter2}

\subsection{Relativistic Mean-field Theory}
In the RMF theory~\cite{ANP1986Serot, RepProgPhys1989Reinhard52, PPNP1996Ring, PhysRep2005Vretenar409, PPNP2006MengJ}, nucleons are described as Dirac spinors moving in a mean potential characterized by the scalar potential $S({\bm r})$ and vector potential $V({\bm r})$. The Dirac equation for a nucleon is,
\begin{equation}
[{\bm \alpha}\cdot{\bm p}+V({\bm r})+\beta(M+S({\bm r}))]\psi_{i}({\bm r})=\varepsilon_{i}\psi_{i}({\bm r}),
\label{EQ:Dirac}
\end{equation}
where ${\bm \alpha}$ and $\beta$ are the Dirac matrices, $M$ is the nucleon mass.
The Dirac equation including the potentials, wave functions, and densities is iteratively solved with the no-sea and the mean-field approximations in either the coordinate space~\cite{NPA1981Horowitz368} or a basis by expansion~\cite{AnnPhys1990Gambhir198, PRC1998Stoitsov58, PRC2003Zhou68}.

For exotic nuclei with large spacial extension, it is not justified to work in the conventional harmonic oscillator (HO) basis. Instead, one can work either in the coordinate space~\cite{NPA1981Horowitz368}, or the improved HO wave function basis~\cite{PRC1998Stoitsov58}, or other bases which have correct asymptotic behaviors such as the Woods-Saxon basis~\cite{PRC2003Zhou68}.

In the coordinate space, one usually solves the Dirac equation~(\ref{EQ:Dirac}) by the shooting method with the box boundary condition and obtains the discretized eigensolutions for the single-particle energy and their corresponding wave functions~\cite{NPA1998MengJ}. If the box is large enough, the shooting method with box boundary condition in the coordinate space is exact for bound states and can describe the exotic nuclei well. However, in this method, the continuum is discretized and there is no information for the widths of resonant states. In order to get the widths for the resonant states, one needs to combine the RMF theory with other methods such as the ACCC approach~\cite{PRC2004ZhangSS70}, the RSM~\cite{PRC2008ZhangL77}, and the CSM~\cite{PRC2010GuoJY82}.

\subsection{Green's Function Method}

The relativistic mean field theory formulated with the Green's function method provides an efficient way to study the single-particle resonate states with correct asymptotic behaviors for the wave functions. The Green's function $\mathcal{G}({\bm r},{\bm r'};\varepsilon)$ describes the propagation of a particle with the energy $\varepsilon$ from ${\bm r}$ to ${\bm r'}$. The relativistic single-particle Green's function, i.e., Green's function for Dirac equation, obeys
\begin{equation}
[\varepsilon-\hat{h}({\bm r})]\mathcal{G}({\bm r},{\bm r'};\varepsilon)=\delta({\bm r}-{\bm r'}),
\end{equation}
where $\hat{h}({\bm r})$ is the Dirac Hamiltonian in Eq.~(\ref{EQ:Dirac}). With a complete set of eigenstates $\psi_{i}({\bm r})$ and eigenvalues $\varepsilon_{i}$ of the Dirac equation,
the relativistic single-particle Green's function can be represented as~\cite{Book2006Eleftherios-GF, PRC2009DaoutidisRing80, PRC2010DingY82}
\begin{equation}
\mathcal{G}({\bm r},{\bm r'};\varepsilon)=\sum_{i}\frac{\psi_{i}({\bm r})\psi_{i}^{\dag}({\bm r'})}{\varepsilon-\varepsilon_{i}},
\label{EQ:GF}
\end{equation}
where $\sum_{i}$ is summation for the discrete states and integral for the continuum explicitly.
Green's function in Eq.~(\ref{EQ:GF}) is analytic on the single-particle complex energy plane with the poles at $\varepsilon_{i}$.

\begin{figure}[!ht]
 \includegraphics[width=0.45\textwidth]{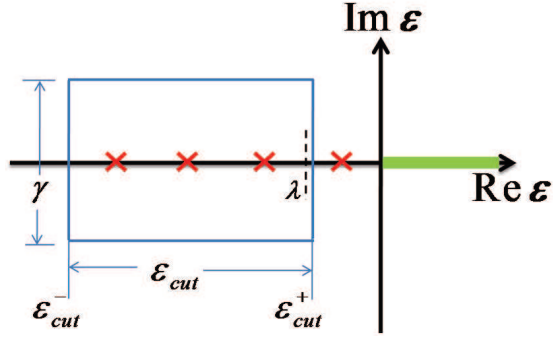}
 \caption{(Color online) Contour path to perform the integrals of the Green's function on the single-particle complex energy plane. The path is chosen to be a rectangle with width $\gamma$ and length $\varepsilon_{\rm cut}$ from $\varepsilon_{\rm cut}^{-}$ to $\varepsilon_{\rm cut}^{+}$. The red crosses denote the discrete single-particle states and the green thick line denotes the continuum. The dashed line is the Fermi surface $\lambda$.}
 \label{Fig1}
\end{figure}

According to Cauchy's theorem, the scalar density $\rho_{s}({\bm r},{\bm r'})$ and vector density $\rho_{v}({\bm r},{\bm r'})$ in the RMF theory can be calculated by the integrals of the Green's function on the single-particle complex energy plane,
\begin{subequations}
\begin{eqnarray}
\rho_{s}({\bm r},{\bm r'})&=& \sum\limits_{i=1}^{A} \bar{\psi}_{i}({\bm r})\psi_{i}({\bm r'}) \nonumber\\
&=&\frac{1}{2\pi i}\oint_{C}d\varepsilon\left[\mathcal{G}^{(11)}({\bm r}, {\bm r'};\varepsilon)-\mathcal{G}^{(22)}({\bm r}, {\bm r'};\varepsilon)\right],\\
\rho_{v}({\bm r},{\bm r'})
&=& \sum\limits_{i=1}^{A}\psi_{i}^{\dag}({\bm r})\psi_{i}({\bm r'})\nonumber\\
&=&\frac{1}{2\pi i}\oint_{C}d\varepsilon\left[\mathcal{G}^{(11)}({\bm r}, {\bm r'};\varepsilon)+\mathcal{G}^{(22)}({\bm r}, {\bm r'};\varepsilon)\right],
\end{eqnarray}
\label{EQ:Density-GF}
\end{subequations}
where $\mathcal{G}^{(11)}({\bm r}, {\bm r'};\varepsilon)$ and $\mathcal{G}^{(22)}({\bm r}, {\bm r'};\varepsilon)$ are respectively the $``11"$ and $``22"$ components of $\mathcal{G}({\bm r}, {\bm r'};\varepsilon)$, $C$ is the contour path for the integral.
Integrating the vector density $\rho_{v}({\bm r},{\bm r})$ over ${\bm r}$ in coordinate space gives the particle number inside the contour path $C$,
\begin{equation}
N=\int d{\bm r}\rho_{v}({\bm r},{\bm r}).
\end{equation}
For a given nucleus, the contour path $C$ in Eq.~(\ref{EQ:Density-GF}) is chosen to enclose all the occupied bound states with energy $\varepsilon_{i}\leq \lambda$~as shown in Fig.~\ref{Fig1}, where the Fermi surface $\lambda$ is determined by the particle number $N$.

With the spherical symmetry, Green's function and densities can be expanded  as
\begin{subequations}
\begin{eqnarray}
\mathcal{G}({\bm r}, {\bm r'};\varepsilon)&=&\sum_{\kappa m}Y_{jm}^{l}(\theta, \phi)\frac{\mathcal{G}_{\kappa}(r,r';\varepsilon)}{rr'}Y_{jm}^{l*}(\theta',\phi'),\\
\rho_{s}({\bm r}, {\bm r'})&=&\sum_{\kappa m}Y_{jm}^{l}(\theta, \phi)\rho_{s,\kappa}(r,r')Y_{jm}^{l*}(\theta',\phi'),\\
\rho_{v}({\bm r}, {\bm r'})&=&\sum_{\kappa m}Y_{jm}^{l}(\theta, \phi)\rho_{v,\kappa}(r,r')Y_{jm}^{l*}(\theta',\phi'),
\end{eqnarray}
\end{subequations}
where $Y_{jm}^{l}(\theta, \phi)$ is the spin spherical harmonic and quantum number $\kappa$ is defined as
$\kappa=(-1)^{j+l+1/2}(j+1/2)$.

According to Eqs.~(\ref{EQ:Density-GF}a) and (\ref{EQ:Density-GF}b), the radial parts of the local scalar density $\rho_{s}({\bm r})=\rho_{s}({\bm r}, {\bm r})$ and vector density $\rho_{v}({\bm r})=\rho_{v}({\bm r}, {\bm r})$ by the Green's function are
\begin{subequations}
\begin{eqnarray}
\rho_{s}(r)&=&\frac{1}{4\pi r^{2}}\frac{1}{2\pi i}\sum_{\kappa}(2j+1)\oint_{C} d\varepsilon \left[\mathcal{G}_{\kappa}^{(11)}(r,r;\varepsilon)-\mathcal{G}_{\kappa}^{(22)}(r,r;\varepsilon)\right],\\
\rho_{v}(r)&=&\frac{1}{4\pi r^{2}}\frac{1}{2\pi i}\sum_{\kappa}(2j+1)\oint_{C} d\varepsilon \left[\mathcal{G}_{\kappa}^{(11)}(r,r;\varepsilon)+\mathcal{G}_{\kappa}^{(22)}(r,r;\varepsilon)\right].
\end{eqnarray}
\label{EQ:Density-rGF}
\end{subequations}

From the densities given by the Green's function, one can calculate the single-particle potentials $V(\bm r)$ and $S(\bm r)$ in Eq.~(\ref{EQ:Dirac}), and the Dirac equations are solved again to provide new Green's functions. In this way, the RMF coupled equations can be solved by iteration self-consistently.

In the RMF-GF method, the energies and widths of single-particle bound and resonate states can be obtained from the density of states $n(\varepsilon)$~\cite{Book2006Eleftherios-GF},
\begin{equation}
n(\varepsilon)=\sum_{i}\delta(\varepsilon-\varepsilon_{i}),
\end{equation}
where $\varepsilon_{i}$ are the eigenvalues of the Dirac equation, $\varepsilon$ is taken along the real-$\varepsilon$ axis, $\sum_{i}$ is
summation for the discrete states and integral for the continuum explicitly, and $n(\varepsilon)d\varepsilon$ gives the number of states in the interval $[\varepsilon, \varepsilon+d\varepsilon]$. For the bound states, the density of states $n(\varepsilon)$ exhibits discrete $\delta$-function at $\varepsilon=\varepsilon_{i}$, while in the continuum $n(\varepsilon)$ has a continuous distribution.

By introducing an infinitesimal imaginary part $``i\epsilon"$ to energy $\varepsilon$, it can be proved that the density of states $n(\varepsilon)$ can be obtained by integrating the imaginary part of the Green's function $\mathcal{G}({\bm r}, {\bm r};\varepsilon+i\epsilon)$ over ${\bm r}$,
\begin{equation}
n(\varepsilon)=-\frac{1}{\pi}\int d{\bm r}{\rm Im}\left[\mathcal{G}^{(11)}({\bm r},{\bm r};\varepsilon+i\epsilon)+\mathcal{G}^{(22)}({\bm r},{\bm r};\varepsilon+i\epsilon)\right].
\label{EQ:DOS}
\end{equation}
With the spherical symmetry, Eq.~(\ref{EQ:DOS}) becomes
\begin{equation}
n(\varepsilon)=\sum_{\kappa}n_{\kappa}(\varepsilon),
\end{equation}
where the density of states for each $\kappa$ is
\begin{equation}
n_{\kappa}(\varepsilon)=-\frac{2j+1}{\pi }\int d{r}{\rm Im}\left[\mathcal{G}_{\kappa}^{(11)}({r},{r};\varepsilon+i\epsilon)+\mathcal{G}_{\kappa}^{(22)}({r},{ r};\varepsilon+i\epsilon)\right].
\label{EQ:DOS-lj}
\end{equation}

For given energy $\varepsilon$ and quantum number $\kappa$, the radial Green's function $\mathcal{G}_{\kappa}(r,r';\varepsilon)$ for the Dirac equation can be constructed as~\cite{PRB1992Tamura45, PRC2009DaoutidisRing80, PRC2010DingY82},
\begin{eqnarray}
&&\mathcal{G}_{\kappa}(r,r';\varepsilon)\nonumber\\&=&\frac{1}{W_{\kappa}(\varepsilon)}
\left[\theta(r-r')\phi_{\kappa}^{(2)}(r,\varepsilon)\phi_{\kappa}^{(1)\dag}
(r',\varepsilon)+\theta(r'-r)\phi_{\kappa}^{(1)}(r,\varepsilon)\phi_{\kappa}^{(2)\dag}
(r',\varepsilon)\right],\label{GFDirac}
\end{eqnarray}
where $\theta(r-r')$ is the step function, $\phi_{\kappa}^{(1)}(r,\varepsilon)$ and $\phi_{\kappa}^{(2)}(r,\varepsilon)$ are two linearly independent Dirac spinors
\begin{equation}
\phi_{\kappa}^{(1)}(r,\varepsilon)=\left(
                                     \begin{array}{c}
                                       g_{\kappa}^{(1)}(r,\varepsilon) \\
                                       f_{\kappa}^{(1)}(r,\varepsilon) \\
                                     \end{array}
                                   \right),~~~~
\phi_{\kappa}^{(2)}(r,\varepsilon)=\left(
                                     \begin{array}{c}
                                       g_{\kappa}^{(2)}(r,\varepsilon) \\
                                       f_{\kappa}^{(2)}(r,\varepsilon) \\
                                     \end{array}
                                   \right),
\end{equation}
obtained from asymptotic behaviors at $r\rightarrow 0$ and $r\rightarrow \infty$ respectively. The $r$-independent $W_{\kappa}(\varepsilon)$ is the Wronskian function defined by
\begin{equation}
W_{\kappa}(\varepsilon)=g_{\kappa}^{(1)}(r, \varepsilon)f_{\kappa}^{(2)}(r, \varepsilon)-g_{\kappa}^{(2)}(r, \varepsilon)f_{\kappa}^{(1)}(r, \varepsilon).
\end{equation}

The Dirac spinor $\phi_{\kappa}^{(1)}(r)$ is regular at the origin and $\phi^{(2)}_{\kappa}(r)$ at $r\rightarrow \infty$ is oscillating outgoing for $\varepsilon>0$ and exponentially decaying for $\varepsilon<0$. Explicitly, Dirac spinor $\phi_{\kappa}^{(1)}(r,\varepsilon)$ at $r\rightarrow 0$ satisfies
\begin{equation}
\phi_{\kappa}^{(1)}(r,\varepsilon)
                                 \rightarrow r\left(
                                                 \begin{array}{c}
                                                   {\displaystyle j_{l}(kr)} \\
                                                  {\displaystyle \frac{\kappa}{|\kappa|}\frac{\varepsilon-V-S}{k}j_{\tilde{l}}(kr)} \\
                                                 \end{array}
                                               \right)
                                 \rightarrow \left(
                                               \begin{array}{c}
                                                 {\displaystyle\frac{r}{(2l+1)!!}(kr)^{l}} \\
                                                 {\displaystyle \frac{\kappa}{|\kappa|}\frac{r(\varepsilon-V-S)}{k(2\tilde{l}+1)!!}(kr)^{\tilde{l}}} \\
                                               \end{array}
                                             \right),
                                             \label{EQ:phi1-r0}
\end{equation}
where $k^{2}=(\varepsilon-V-S)(\varepsilon-V-S+2M)>0$, quantum number $\tilde{l}$ is defined as $\tilde{l}=l+(-1)^{j+l+1/2 }$, and $j_{l}(kr)$ is the spherical Bessel function of the first kind~\cite{Book1990Greiner-RQM}.

The Dirac spinor $\phi_{\kappa}^{(2)}(r,\varepsilon)$ at $r\rightarrow \infty$ satisfies
\begin{equation}
\phi_{\kappa}^{(2)}(r,\varepsilon)
                                 \rightarrow \left(
                                               \begin{array}{c}
                                                 rh_{l}^{(1)}(kr) \\
                                                 {\displaystyle \frac{\kappa}{|\kappa|}}
                                                 {\displaystyle \frac{ikr}{\varepsilon+2M}h_{\tilde{l}}^{(1)}(kr)} \\
                                               \end{array}
                                             \right)
                                 \rightarrow \left(
                                               \begin{array}{c}
                                                 {\displaystyle 1} \\
                                                 {\displaystyle \frac{\kappa}{|\kappa|}\frac{ik}{\varepsilon+2M}} \\
                                               \end{array}
                                             \right) {\displaystyle e^{ikr}},
                                             \label{EQ:phi2-rmax-1}
\end{equation}
for $\varepsilon>0$ and
 \begin{equation}
 \phi_{\kappa}^{(2)}(r,\varepsilon)
                                 \rightarrow \left(
                                               \begin{array}{c}
                                                 {\displaystyle r\sqrt{\frac{2Kr}{\pi}}K_{l+\frac{1}{2}}(Kr)} \\
                                                {\displaystyle  \frac{-Kr}{\varepsilon+2M}\sqrt{\frac{2Kr}{\pi}}K_{\tilde{l}+\frac{1}{2}}(Kr)} \\
                                               \end{array}
                                             \right)
                                 \rightarrow \left(
                                               \begin{array}{c}
                                                {\displaystyle  1 }\\
                                                {\displaystyle -\frac{K}{\varepsilon+2M} }\\
                                               \end{array}
                                             \right){\displaystyle e^{-Kr}},
                                             \label{EQ:phi2-rmax-2}
 \end{equation}
 for $\varepsilon<0$. Here, $K^{2}=(V-S-\varepsilon)(\varepsilon-V+S+2M)>0$, $h_{l}^{(1)}(kr)$ is the spherical Hankel function of the first kind, and $K_{l+\frac{1}{2}}(Kr)$ the modified spherical Bessel function~\cite{Book1970Abramowitz-HandbookMathFunction}.

\section{NUMERICAL DETAILS}\label{Chapter3}

Taking the nucleus $^{120}{\rm Sn}$ as an example, the energies and widths for the single-neutron resonant states are investigated by the RMF-GF method with effective interactions PK1~\cite{PRC2004LongWH69} and NL3~\cite{PRC1997Lalazissis55}. The obtained results are compared with those by the shooting method with box boundary condition for the bound states and those from the RMF-S method~\cite{PRC2002CaoLG66}, the RMF-ACCC approach~\cite{PRC2004ZhangSS70}, the RMF-RSM~\cite{PRC2008ZhangL77}, and the RMF-CSM~\cite{PRC2010GuoJY82} for the resonant states.

In order to construct the radial Dirac Green's function of Eq.~(\ref{GFDirac}) in the coordinate space, Runge-Kutta algorithm with space size $R_{\rm max}$ and step $dr=0.1~{\rm fm}$ is used to obtain the two independent solutions $\phi_{\kappa}^{(1)}(r)$ and $\phi_{\kappa}^{(2)}(r)$  from asymptotic behaviors of Dirac spinor at $r\rightarrow 0$ (Eq.~(\ref{EQ:phi1-r0})) and $r\rightarrow \infty$ (Eqs.~(\ref{EQ:phi2-rmax-1}) and (\ref{EQ:phi2-rmax-2})) respectively. To perform the integrals of the Green's function in Eq.~(\ref{EQ:Density-rGF}), the contour path $C$ is chosen to be a rectangle on the single-particle complex energy plane as shown in Fig.~\ref{Fig1}. The width $\gamma$ is taken as $0.1~{\rm MeV}$. To enclose all the occupied single-particle levels, the path starts from the bottom of the mean potential $e_{\rm cut}^{-} \sim V(0)+S(0)$ and ends around the particle Fermi surface $e_{\rm cut}^{+} \sim \lambda$. The energy step is taken as $d\varepsilon=0.005~{\rm MeV}$ on the contour path. To calculate the density of states $n_{\kappa}(\varepsilon)$ along the real-$\varepsilon$ axis, the parameter $\epsilon$ in Eq.~(\ref{EQ:DOS-lj}) is taken as $1\times10^{-6}~{\rm MeV}$ and the energy step along the real-$\varepsilon$ axis is $1\times10^{-4}~{\rm MeV}$.

\section{RESULTS AND DISCUSSION}\label{Chapter4}

Taking $^{120}$Sn as an example, we will study the energies of single-neutron bound states and the energies and widths of single-neutron resonant states from the density of neutron states $n_{\kappa}(\varepsilon)$.

\begin{figure}[!ht]
 \includegraphics[width=0.45\textwidth]{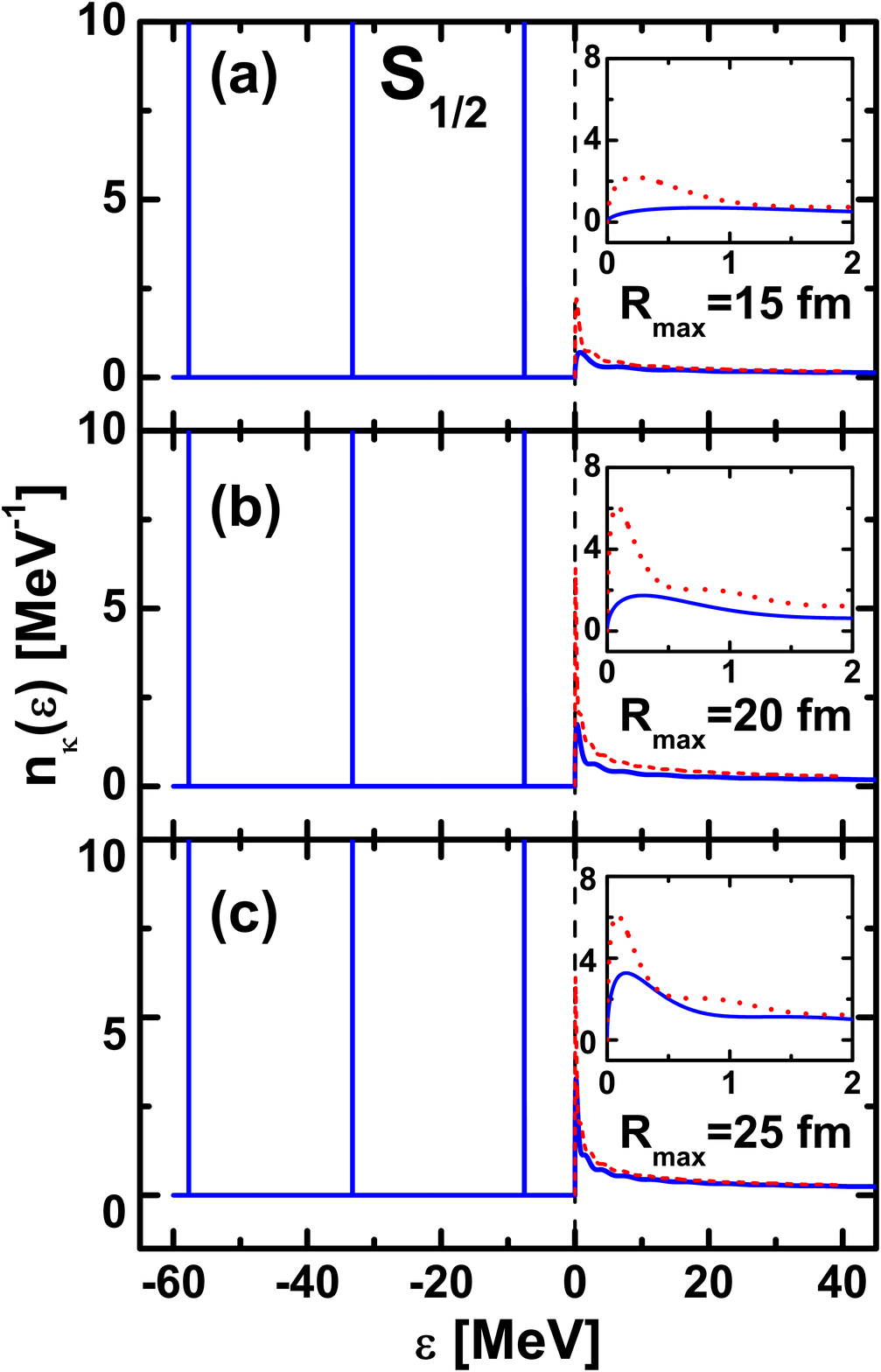}
  \caption{(Color online) Density of neutron states $n_{\kappa}(\varepsilon)$ for $s_{1/2}$ block in $^{120}{\rm Sn}$ obtained by the RMF-GF method with PK1 and space sizes $R_{\rm max}=15~{\rm fm}$(a), $20~{\rm fm}$(b), $25~{\rm fm}$(c) respectively. Spectra with $\varepsilon>0$ are compared with $n_{\kappa}(\varepsilon)$ obtained with potentials $V=S=0$ (denoted by the short-dashed line). The dashed line is the continuum threshold. The inserts enlarge the density of neutron states in the energy range $\varepsilon \in [0,2]~{\rm MeV}$. }
 \label{Fig2}
 \end{figure}

Figure~\ref{Fig2} shows the density of neutron states $n_{\kappa}(\varepsilon)$ for $s_{1/2}$ block in $^{120}{\rm Sn}$ obtained by the RMF-GF method with PK1 and $R_{\rm max}=15~{\rm fm}$(a), $20~{\rm fm}$(b), $25~{\rm fm}$(c) respectively. Below the continuum threshold, three peaks of $\delta$-functional shape are observed, which respectively correspond to three bound states, i.e., $1s_{1/2}$, $2s_{1/2}$, and $3s_{1/2}$. Spectra with $\varepsilon>0$ are continuous with a peak close to zero and changing with $R_{\rm max}$. To examine the detailed structures in continuum, the inserts in Fig.~\ref{Fig2} show the density of neutron states for $\varepsilon \in [0,2]~{\rm MeV}$. It can be seen that there appears no peak for $R_{\rm max}=15~{\rm fm}$. For $R_{\rm max}=20~{\rm fm}$, a wide hump appears around $0.30~{\rm MeV}$, which moves to $0.15~{\rm MeV}$ for $R_{\rm max}=25~{\rm fm}$. It is known that the energy of a resonant state is constant against the sizes of the basis or the box. Combining with the discussions in the following, the existence of a single-neutron $s_{1/2}$ resonant state in $^{120}$Sn can be excluded.

In Fig.~\ref{Fig2}, the densities of neutron states for $s_{1/2}$ block in $^{120}{\rm Sn}$ in the continuum are compared with those obtained with potentials $V=S=0$ (denoted by the short-dashed line). The centrifugal potential for the $s_{1/2}$ block is zero, and thus $s$ waves are free wave functions for $V=S=0$. For a free particle with mass $m$ moving in one dimension space $[0, L]$, its density of states can be expressed as a function of energy $\varepsilon$ as $\displaystyle {n(\varepsilon)=\frac{L}{2\sqrt{2}\pi}\sqrt{\frac{m}{\varepsilon}}}$~\cite{Book2008WangZC}, which suggests a divergence with $\varepsilon\rightarrow 0$. From Fig.~\ref{Fig2}, it can be seen that, similar to $n_{\kappa}(\varepsilon)$ for $^{120}{\rm Sn}$, the density of states obtained with $V=S=0$ ( short-dashed line) also shows a peak close to threshold, which changes with $R_{\rm max}$. This suggests that the peak of $n_{\kappa}(\varepsilon)$  close to threshold for $s_{1/2}$ block in $^{120}{\rm Sn}$ comes from the non-resonant continuum, similar to a free particle without confining potential. The different heights of the peaks for $^{120}{\rm Sn}$ and $V=S=0$ are due to the different depths of the potentials.

\begin{figure}[!ht]
 \includegraphics[width=0.45\textwidth]{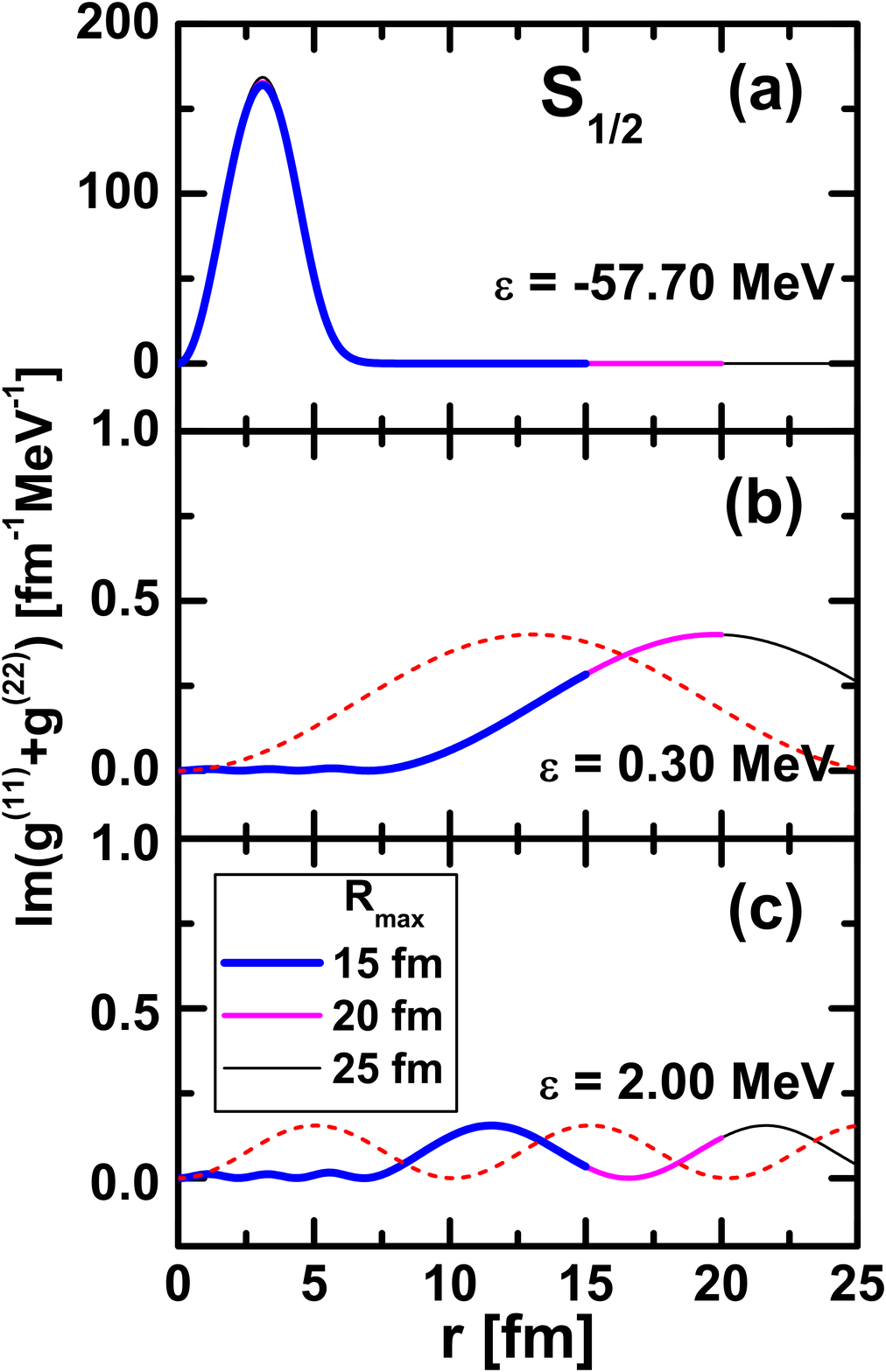}
 \caption{(Color online)  The integrands for the density of states, ${\rm Im}[\mathcal{G}_{\kappa}^{(11)}({r},{r};\varepsilon+i\epsilon)+\mathcal{G}_{\kappa}^{(22)}({r},{ r};\varepsilon+i\epsilon)]$, in Eq.~(\ref{EQ:DOS-lj}) at energy $\varepsilon=-57.70$(a), $0.30$(b), and $2.00~{\rm MeV}$(c) for $s_{1/2}$ block in $^{120}{\rm Sn}$. Calculations are done with $R_{\rm max}=15$, $20$, and $25~{\rm fm}$. For comparison, integrands for $n_{\kappa}(\varepsilon)$ obtained with $V=S=0$ and $R_{\rm max}=25~{\rm fm}$ are shown as the short-dashed lines.}
 \label{Fig3}
\end{figure}

To further understand the density of neutron states for $s_{1/2}$ block in $^{120}{\rm Sn}$, we show in Fig.~\ref{Fig3} the integrands in Eq.~(\ref{EQ:DOS-lj}), i.e., ${\rm Im }[\mathcal{G}_{\kappa}^{(11)}({r},{r};\varepsilon+i\epsilon)+\mathcal{G}_{\kappa}^{(22)}({r},{ r};\varepsilon+i\epsilon)]$, for $\varepsilon = -57.70~{\rm MeV}$(a), $0.30~{\rm MeV}$(b) and $2.00~{\rm MeV}$(c), which respectively correspond to the bound state $1s_{1/2}$, the peak of $n_{\kappa}(\varepsilon)$ in the continuum in Fig.~\ref{Fig2}(b), and an arbitrary energy in the continuum. Calculations are done with $R_{\rm max}=15$, $20$, and $25~{\rm fm}$. For comparison, the integrand obtained with $V=S=0$ and $R_{\rm max}=25~{\rm fm}$ is also shown as the short-dashed lines. The integrand ${\rm Im}[\mathcal{G}_{\kappa}^{(11)}({r},{r};\varepsilon+i\epsilon)+\mathcal{G}_{\kappa}^{(22)}({r},{ r};\varepsilon+i\epsilon)]$
, which corresponds to the vector density of Eq.~(\ref{EQ:Density-rGF}b) at energy $\varepsilon$, is calculated from the single-particle wave functions at energy $\varepsilon$ by Eq.~(\ref{GFDirac}). The integrand at $-57.70~{\rm MeV}$ in Fig.~\ref{Fig3}(a) is peaked around $r\sim 3~{\rm fm}$ and quickly decreases to zero for $r > 7 ~{\rm fm}$, exhibiting the behavior of a bound state. In contrast, the dominant parts of the integrands at $0.30~{\rm MeV}$ in Fig.~\ref{Fig3}(b) and $2.00~{\rm MeV}$ in Fig.~\ref{Fig3}(c) are located in the region $r>7~{\rm fm}$ and oscillating, similar to those short-dashed lines for the free particles obtained with $V=S=0$. In panels (b) and (c), the small amplitudes in the region $r <7~{\rm fm}$ and the different phases of the integrands for $^{120}{\rm Sn}$ compared with those for $V=S=0$ originate from the attractive potential~\cite{Book1972Taylor-ScatteringTheor}, which leads to the peak differences near the threshold in the density of states for $^{120}{\rm Sn}$ and $V=S=0$ in Fig.~\ref{Fig2}.
It is noted that, for a given energy $\varepsilon$, the integrands are the same for different $R_{\rm max}$, which demonstrates that RMF-GF method properly treats the asymptotic behaviors of single-particle wave functions.

\begin{figure}[!ht]
 \includegraphics[width=0.45\textwidth]{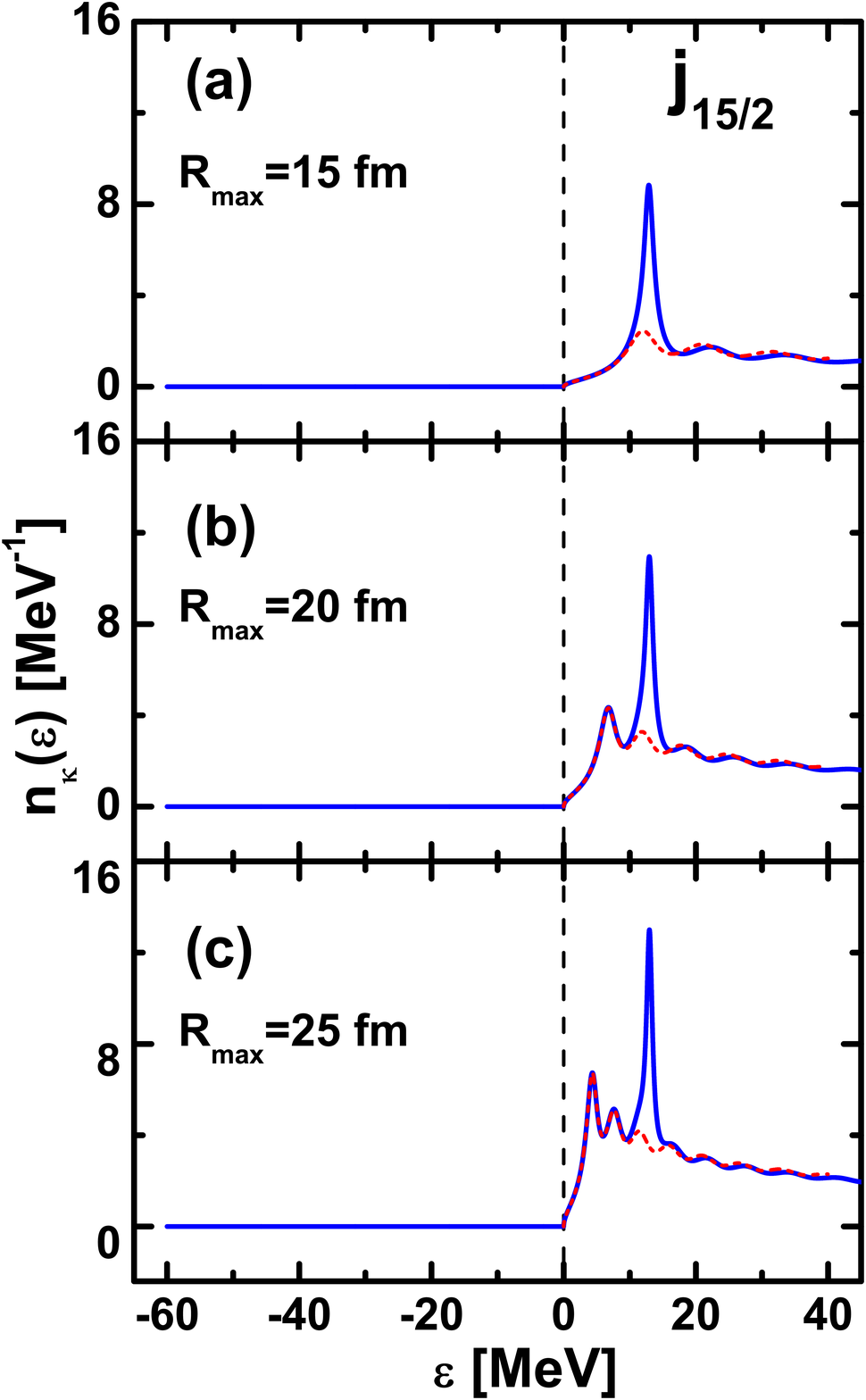}
 \caption{(Color online) The same as Fig.~\ref{Fig2}, but for $j_{15/2}$ block.}
 \label{Fig4}
\end{figure}

In Fig.~\ref{Fig4}, similar to $s_{1/2}$ block in Fig.~\ref{Fig2}, the densities of neutron states $n_{\kappa}(\varepsilon)$ for $j_{15/2}$ block in $^{120}$Sn are shown with PK1 and $R_{\rm max}=15$(a), $20$(b), $25~{\rm fm}$(c) respectively. For comparison, the results obtained with $V=S=0$ are also plotted (denoted by the short-dashed line). In all the three panels, a peak around $12.95~{\rm MeV}$ independent on $R_{\rm max}$ is observed, which demonstrates itself as a resonant state. Apart from this resonant state, the other peaks depend on $R_{\rm max}$ and coincide with the peaks of $n_{\kappa}(\varepsilon)$ obtained with $V=S=0$, which demonstrates their non-resonant characters.

\begin{figure}[!ht]
 \includegraphics[width=0.45\textwidth]{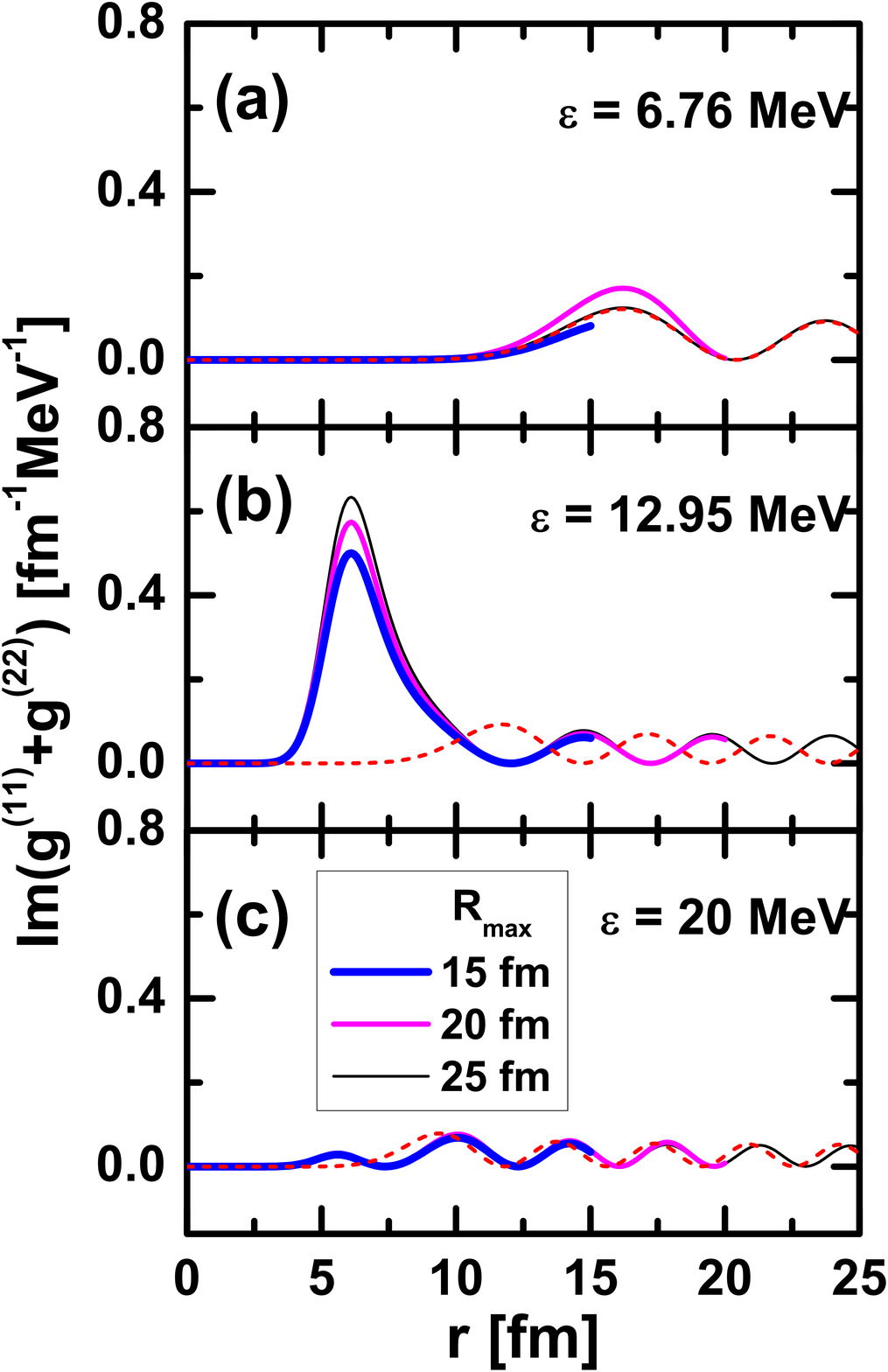}
 \caption{(Color online) The same as Fig.~\ref{Fig3}, but for $j_{15/2}$ block. }
 \label{Fig5}
\end{figure}

In Fig.~\ref{Fig5}, similar to $s_{1/2}$ block in Fig.~\ref{Fig3}, the integrands  ${\rm Im}[\mathcal{G}_{\kappa}^{(11)}({r},{r};\varepsilon+i\epsilon)+\mathcal{G}_{\kappa}^{(22)}({r},{ r};\varepsilon+i\epsilon)]$
for $j_{15/2}$ block are shown respectively for $\varepsilon= 6.76$(a), $12.95$(b), and $20~{\rm MeV}$(c). The solid lines denote the results for $^{120}{\rm Sn}$ with PK1 and $R_{\rm max}=15$, $20$, and $25~{\rm fm}$ and the short-dashed lines correspond to those obtained with $V=S=0$ and $R_{\rm max}=25~{\rm fm}$. The energies $6.76~{\rm MeV}$ and $12.95~{\rm MeV}$ respectively correspond to the first two peaks of $n_{\kappa}(\varepsilon)$ with $R_{\rm max}=20~{\rm fm}$ in Fig.~\ref{Fig4}(b), and $\varepsilon=20~{\rm MeV}$ an arbitrary energy in the continuum. In Fig.~\ref{Fig5}(b), similar to the bound state $1s_{1/2}$ in Fig.~\ref{Fig3}(a), the integrand for $j_{15/2}$ block at $\varepsilon=12.95~{\rm MeV}$ is peaked around $r \sim 6~{\rm fm}$ and quickly decreases, which demonstrates its resonant character. The oscillating tail of the integrands in Fig.~\ref{Fig5}(b) indicates its mixture with the non-resonant continuum. In Fig.~\ref{Fig5}(a) and Fig.~\ref{Fig5}(c), the integrands at $\varepsilon=6.76~{\rm MeV}$ and $\varepsilon=20~{\rm MeV}$ are located mainly outside the nuclear surface. Similar to the discussions on the integrands for $s_{1/2}$ block in Fig.~\ref{Fig3}, by comparing the integrands for $j_{15/2}$ block in $^{120}{\rm Sn}$ with those obtained with $V=S=0$, we can conclude that the spectra at $\varepsilon=6.76$ and $20~{\rm MeV}$ are non-resonant continuum.

\begin{figure}[!ht]
 \includegraphics[width=0.9\textwidth]{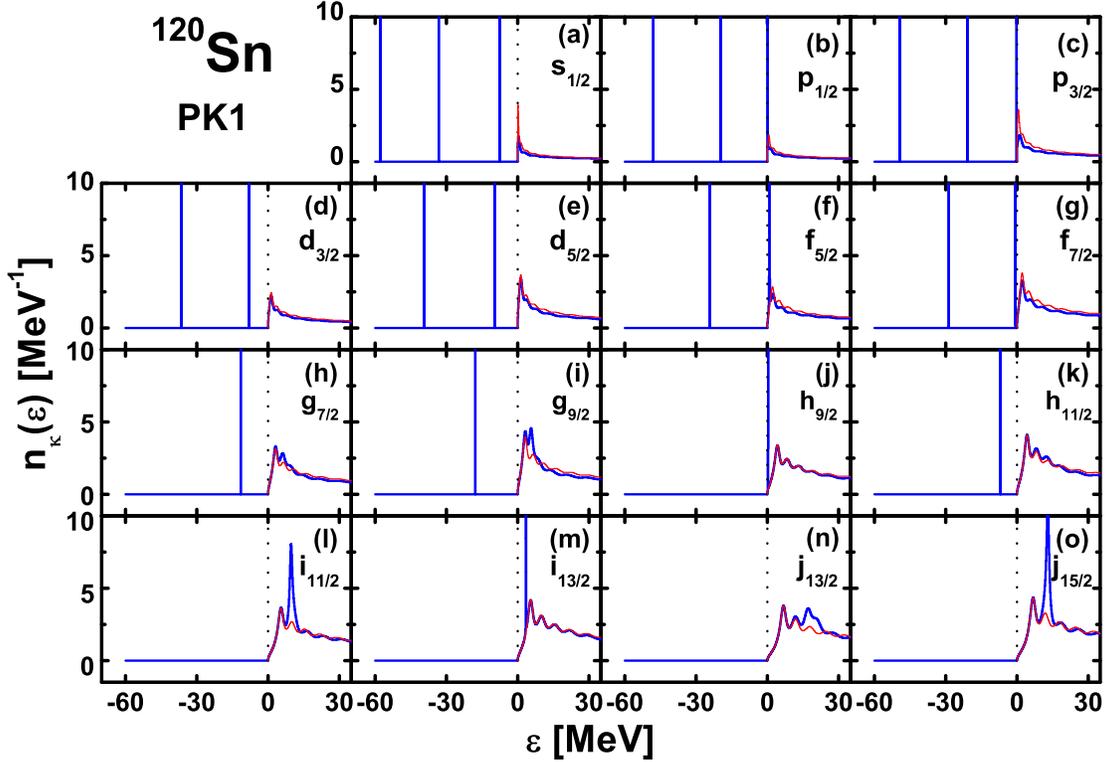}
 \caption{(Color online) Density of neutron states $n_{\kappa}(\varepsilon)$ for different blocks in $^{120}{\rm Sn}$ calculated by RMF-GF method with PK1 and $R_{\rm max}=20~{\rm fm}$ (blue solid line) and compared with $n_{\kappa}(\varepsilon)$ obtained with $V=S=0$ (red dotted line). The dashed line represents the continuum threshold.}
 \label{Fig6}
\end{figure}

In Fig.~\ref{Fig6}, the densities of neutron states $n_{\kappa}(\varepsilon)$ for different blocks in $^{120}{\rm Sn}$ obtained by the RMF-GF method with PK1 and $R_{\rm max}=20~{\rm fm}$ are summarized and compared with those obtained with $V=S=0$.
The single-neutron bound states are observed in $s_{1/2}$, $p_{1/2}$, $p_{3/2}$, $d_{3/2}$, $d_{5/2}$, $f_{5/2}$, $f_{7/2}$, $g_{7/2}$, $h_{9/2}$, and $h_{11/2}$ blocks. The resonant states are observed in $p_{1/2}$, $f_{5/2}$, $g_{9/2}$, $i_{11/2}$, $i_{13/2}$, $j_{13/2}$, and $j_{15/2}$ blocks, in which there are peaks of $n_{\kappa}(\varepsilon)$ on top of those obtained with $V=S=0$. Furthermore, the peak energies of these resonant states are independent on the $R_{\rm max}$.

\begin{table}[!ht]
\center
\caption{Single-neutron energies for the bound states in $^{120}{\rm Sn}$ extracted from $n_{\kappa}(\varepsilon)$ in Fig.~\ref{Fig6} by RMF-GF method with PK1 and $R_{\rm max}=20~{\rm fm}$, in comparison with those obtained by the shooting method with the box boundary condition. $|\Delta\varepsilon|$ represents the difference between the results by the two methods. All quantities are in MeV.}
\label{Tab1}
\begin{tabular}{ccccccccc}
  \hline\hline
  \multirow{2}*{$nl_{j}$ }   & \multicolumn{3}{c}{positive parity } && \multirow{2}*{$nl_{j}$ } &  \multicolumn{3}{c}{negative parity }\\
           \cline{2-4}\cline{7-9}
      &~~$\varepsilon_{\rm GF}~~$&$~~\varepsilon_{\rm box}~~$&$|\Delta\varepsilon|$&~~~~&&$~~\varepsilon_{\rm GF}~~$& $~~\varepsilon_{\rm box}~~$&$|\Delta\varepsilon|$\\\hline
  $1s_{1/2}$&-57.7043 &-57.7043&0.0000 &   &$1p_{3/2}$ &-49.2053&-49.2053&0.0000\\
  $1d_{5/2}$&-39.3590 &-39.3590&0.0000 &   &$1p_{1/2}$ &-47.9166&-47.9166&0.0000\\
  $1d_{3/2}$&-36.5029 &-36.5029&0.0000 &   &$1f_{7/2}$ &-28.7538&-28.7538&0.0000\\
  $2s_{1/2}$&-33.2498 &-33.2498&0.0000 &   &$1f_{5/2}$ &-24.0787&-24.0787&0.0000\\
  $1g_{9/2}$&-17.8306 &-17.8306&0.0000 &   &$2p_{3/2}$ &-20.9125&-20.9125&0.0000\\
  $1g_{7/2}$&-11.4990 &-11.4990&0.0000 &   &$2p_{1/2}$ &-19.5882&-19.5882&0.0000\\
  $2d_{5/2}$&~-9.7489 &~-9.7489&0.0000 &   &$1h_{11/2}$&~-6.9592&~-6.9592&0.0000\\
  $2d_{3/2}$&~-7.9307 &~-7.9307&0.0000 &   &$2f_{7/2}$ &~-0.5708&~-0.5692&0.0016\\
  $3s_{1/2}$&~-7.5663 &~-7.5663&0.0000 &   &$3p_{3/2}$ &~-0.1371&~-0.0820&0.0551\\
  \hline\hline
\end{tabular}
\end{table}

In Table~\ref{Tab1}, single-neutron energies for the bound states in $^{120}{\rm Sn}$ extracted from the density of states $n_{\kappa}(\varepsilon)$ by the RMF-GF method with PK1 and $R_{\rm max}=20~{\rm fm}$ in Fig.~\ref{Fig6} (labeled as $\varepsilon_{\rm GF}$) are listed, in comparison with those obtained by the shooting method with box boundary condition (labeled as $\varepsilon_{\rm box}$). From the energy difference $|\Delta\varepsilon|$ between $\varepsilon_{\rm GF}$ and $\varepsilon_{\rm box}$ in Table~\ref{Tab1}, it can be seen that same energies are obtained for the deeply bound levels and differences exist for the very weakly bound levels $2f_{7/2}$ and $3p_{3/2}$, indicating that large box size is necessary for the weakly bound states.

\begin{table}[!ht]
\center
\caption{Energies and widths $\varepsilon(\Gamma)$ of single-neutron resonant states in $^{120}{\rm Sn}$ obtained by different methods. RMF-GF, RMF-S~\cite{PRC2002CaoLG66}, RMF-RSM~\cite{PRC2008ZhangL77}, RMF-CSM~\cite{PRC2010GuoJY82}, and RMF-ACCC~\cite{PRC2004ZhangSS70} respectively represent results from the Green's function method, the scattering phase-shift method, the real stabilization method, the complex stabilization method, and the analytical continuation in the coupling constant approach based on the RMF theory with PK1 and/or NL3. All quantities are in MeV.}
\label{Tab2}
\begin{tabular}{clcccccc}
  \hline\hline
&             &$p_{1/2}$   &$h_{9/2}$   &$f_{5/2}$   &$i_{13/2}$  &$i_{11/2}$  &$j_{15/2}$ \\\hline
\multirow{2}*{PK1}&RMF-GF  &0.031(0.086)&0.251($<0.001$)&0.887(0.064)&3.469(0.003)&9.700(1.272) &12.956(1.375)\\
&RMF-RSM &            &               &0.870(0.064)&3.469(0.005)&9.811(1.275) &12.865(1.027)\\\hline
&RMF-GF  &0.017(0.109)&0.232($<0.001$)  &0.685(0.042)&3.264(0.003)&9.465(1.214) &12.588(1.340)\\
&RMF-S   &0.176(0.316)&0.229(0.000)   &0.657(0.031)&3.261(0.004)&9.751(1.384) &12.658(1.051)\\
NL3&RMF-RSM &            &               &0.674(0.030)&3.266(0.004)&9.559(1.205) &12.564(0.973)\\
&RMF-CSM &            &               &0.670(0.020)&3.266(0.004)&9.597(1.212) &12.578(0.992)\\
&RMF-ACCC&0.072(0.000)&0.232(0.000)   &0.685(0.023)&3.262(0.004)&9.600(1.110) &12.600(0.900)\\
  \hline\hline
\end{tabular}
\end{table}

From the density of states, one can also extract the energies $\varepsilon$ and widths $\Gamma$ of single-particle resonant states. Here, the width $\Gamma$ is defined as the full-width at half-maximum (FWHM) of peaks. The background of the non-resonant continuum in the density of states, i.e., that obtained with $V=S=0$, is removed when reading the widths of resonant states. In Table~\ref{Tab2}, we list the energies $\varepsilon$ and widths $\Gamma$ of single-neutron resonant states in $^{120}{\rm Sn}$ extracted from $n_{\kappa}(\varepsilon)$ in Fig.~\ref{Fig6}, in comparison with the previous results by the RMF-RSM~\cite{PRC2008ZhangL77}. It can be seen that the RMF-GF calculations provide an excellent agreement with the RMF-RSM method for the single-neutron resonant states. In Table~\ref{Tab2}, we also list the results by RMF-GF calculations performed with NL3 and $R_{\rm max}=20~{\rm fm}$, and those from the RMF-S method~\cite{PRC2002CaoLG66}, RMF-RSM~\cite{PRC2008ZhangL77}, RMF-CSM~\cite{PRC2010GuoJY82}, and the RMF-ACCC approach~\cite{PRC2004ZhangSS70}. Similar to the results for PK1, the ones for NL3 by different methods provide remarkable consistency in the energies and the widths of the single-neutron resonant states in $^{120}{\rm Sn}$.

\begin{figure}[!ht]
 \includegraphics[width=0.45\textwidth]{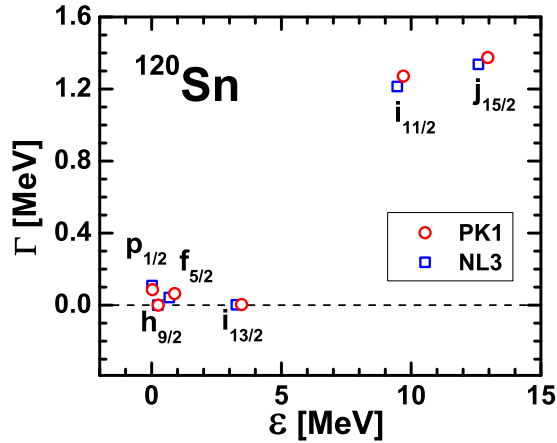}
 \caption{Energies $\varepsilon$ and widths $\Gamma$ of single-neutron resonant states $p_{1/2}$, $h_{9/2}$, $f_{5/2}$, $i_{13/2}, i_{11/2}$, and $j_{15/2}$ in $^{120}{\rm Sn}$ obtained by the Green's function method in the framework of RMF theory with PK1 and NL3. }
 \label{Fig7}
\end{figure}

In Fig.~\ref{Fig7}, the single-neutron resonant states in $^{120}{\rm Sn}$ calculated by the Green's function method in the framework of RMF theory with effective interactions PK1 and NL3 are given in the energy $\varepsilon$ and width $\Gamma$ plane.

 In the RMF-GF method, the single-particle bound states and continuum are treated on the same footing. Taking the  density of states for free particle as a reference, the energies and widths of single-particle resonant states are extracted from the density of states without any ambiguity. The results demonstrate that the Green's function method is suitable and reliable in describing single-neutron resonant states.

\section{SUMMARY}\label{Chapter5}

In summary, the RMF theory formulated with the Green's function method in coordinate space is developed to investigate the single-particle resonant states. Taking the density of states for free particle as a reference, the energies and widths of single-particle resonant states are extracted from the density of states without any ambiguity. As an example, the obtained energies and widths for the single-neutron resonant states in nucleus $^{120}$Sn are compared with those by the scattering phase-shift method, the analytic continuation in the coupling constant approach, the real stabilization method, and the complex scaling method. Excellent agreements are found for the energies and widths of single-neutron resonant states by these methods.

\begin{acknowledgments}
Helpful discussions with Nguyen Van Giai, Jian-You Guo, Bing-Nan Lu, Masayuki Matsuo, and Shan-Gui Zhou are acknowledged. This work was supported in part by the Major State 973 Program of China (Grant No. 2013CB834400), the National Natural Science Foundation of China (Grants No. 11175002,
No. 11335002, No. 11375015, No. 11345004, No. 11405116, No. 11405090), Research Fund for the Doctoral Program of Higher Education (Grant No. 20110001110087).
\end{acknowledgments}
%
%\bibliography{bibliography}

\end{CJK*}
\end{document}